\documentclass[aps,amsmath,notitlepage,twocolumn,amssymb,prl,longbibliography]{revtex4-1}
\usepackage{graphicx}
\usepackage{dcolumn}
\usepackage{bm}
\usepackage[colorlinks=true,linkcolor = blue,citecolor=blue,urlcolor=blue,anchorcolor = blue]{hyperref}
\usepackage{wasysym}
\usepackage{stmaryrd}
\usepackage{verbatim}
\usepackage{subfigure}
\usepackage{amsmath}
\usepackage{times}
\usepackage[version=4]{mhchem}
\usepackage{braket}
\usepackage{siunitx}
\usepackage{amssymb}
\usepackage{color}
\usepackage{float}
\newcommand{\RNum}[1]{\uppercase\expandafter{\romannumeral #1\relax}}

\begin{document}

\title{Ground State Magnetic Structure and Magnetic Field Effects in the Layered Honeycomb Antiferromagnet \ce{YbOCl}}
\author{Zheng\,Zhang$^{1}$}
\author{Yanzhen\,Cai$^{2}$}
\author{Jinlong\,Jiao$^{3}$}
\author{Jing\,Kang$^{1}$}
\author{Dehong\,Yu$^{4}$}
\author{Bertrand Roessli$^{5}$}
\author{Anmin\,Zhang$^{2}$}
\author{Jianting\,Ji$^{1}$}
\author{Feng\,Jin$^{1}$}
\author{Jie\,Ma$^{3}$}
\email[e-mail:]{jma3@sjtu.edu.cn}
\author{Qingming\,Zhang$^{1,2}$}
\email[e-mail:]{qmzhang@iphy.ac.cn}

\affiliation{$^{1}$Beijing National Laboratory for Condensed Matter Physics, Institute of Physics, Chinese Academy of Sciences, Beijing 100190, China}
\affiliation{$^{2}$School of Physical Science and Technology, Lanzhou University, Lanzhou 730000, China}
\affiliation{$^{3}$Department of Physics and Astronomy, Shanghai Jiao Tong University, Shanghai 200240, China}
\affiliation{$^{4}$Australian Nuclear Science and Technology Organisation, Lucas Heights, New South Wales 2232, Australia}
\affiliation{$^{5}$Laboratory for Neutron Scattering and Imaging, Paul Scherrer Institut, 5232 Villigen, Switzerland}

\begin{abstract}

\ce{YbOCl} is a representative member of the van der Waals layered honeycomb rare-earth chalcohalide REChX (RE = rare earth, Ch = O, S, Se, and Te, and X = F, Cl, Br, and I) family reported recently. Its spin ground state remains to be explored experimentally.
In this paper, we have grown high-quality single crystals of \ce{YbOCl} and conducted comprehensive thermodynamic, elastic, and inelastic neutron scattering experiments down to 50 mK. 
The experiments reveal an antiferromagnetic phase below 1.3 K, which is identified as a spin ground state with an intralayer ferromagnetic and interlayer antiferromagnetic ordering.
By applying sophisticated numerical techniques to a honeycomb (nearest-neighbor)-triangle (next-nearest-neighbor) model Hamiltonian which accurately describes the highly anisotropic spin system, we are able to well simulate the experiments and determine the diagonal and off-diagonal spin-exchange interactions. The simulations give an antiferromagnetic Kitaev term comparable to the Heisenberg one. 
The experiments under magnetic fields allow us to establish a magnetic field-temperature phase diagram around the spin ground state. 
Most interestingly, a relatively small magnetic field ($\sim$ 0.3 to 3 T) can significantly suppress the antiferromagnetic order, suggesting an intriguing interplay of the Kitaev interaction and magnetic fields in the spin system. 
The present study provides fundamental insights into the highly anisotropic spin systems and opens a new window to look into Kitaev spin physics in a rare-earth-based system.

\end{abstract}

\maketitle

\subsection{Introduction}
Among the various proposed quantum spin liquid (QSL) models, the exactly solvable Kitaev spin liquid (KSL) model built on a honeycomb lattice, has garnered significant interest due to its fundamental importance and potential applications in topological quantum computing. However, the material realization of KSL remains a crucial challenge, owing to the rather unusual bond-dependent spin interactions. In recent years, the efforts in searching for the KSL candidates have been focused on 4d/5d ions, such as iridate \ce{A2IrO3} (A = Li, Na, and Cu)\cite{PhysRevLett.114.077202,PhysRevB.83.220403,PhysRevB.82.064412,PhysRevLett.110.097204,Abramchuk2017,PhysRevLett.122.167202} and $\alpha$-\ce{RuCl3}\cite{PhysRevLett.119.037201,PhysRevLett.119.227208,PhysRevLett.120.217205,PhysRevLett.119.227202}.
In general, rare-earth magnetic ions exhibit highly anisotropic magnetism stemming from the strong spin-orbit coupling (SOC), and can naturally serve as building blocks for Kitaev materials through the Jackeli-Khaliullin mechanism\cite{PhysRevLett.102.017205}. Particularly, rare-earth ions with an odd number of 4f electrons possess a doubly degenerate crystalline electric field (CEF) ground state (Kramers doublets), which is protected by time-reversal symmetry and yields an effective spin-1/2 required by KSL. Due to the critical experimental requirements, Kitaev spin systems based on rare-earth magnetic ions remain less explored\cite{PhysRevB.100.180406,PhysRevB.102.014427}.

Recently, we have revealed a family of KSL candidates, namely rare-earth chalcohalides REChX (RE = rare-earth; Ch = O, S, Se, and Te; X = F, Cl, Br, and I)\cite{JiantingJi,PhysRevResearch.4.033006}. Most of the family members possess a high symmetry of R-3m, with nearest rare-earth magnetic ions forming an undistorted honeycomb lattice. Furthermore, the van der Waals-layered rare-earth chalcohalides offer a good two-dimensionality, which is crucial for Abelian/non-Abelian excitations in KSLs and conducive to the fundamental study and potential applications based on atomically thin flakes. 

Following the initial discovery, we have successfully grown sizable single crystals of \ce{YbOCl} ($\sim$ 15 mm in maximum), providing an exceptional opportunity for in-depth experimental studies on the spin systems. In our previous work, we have conducted thermodynamic measurements above 1.8 K, examined CEF excitations, and initially studied the theoretical ground state phase diagram\cite{PhysRevResearch.4.033006}. The experimental investigation of the spin ground state, which is a fundamental issue in the search for the KSL phase, remains unexplored in the spin system so far.

\begin{figure*}[t]
	\includegraphics[scale=0.9]{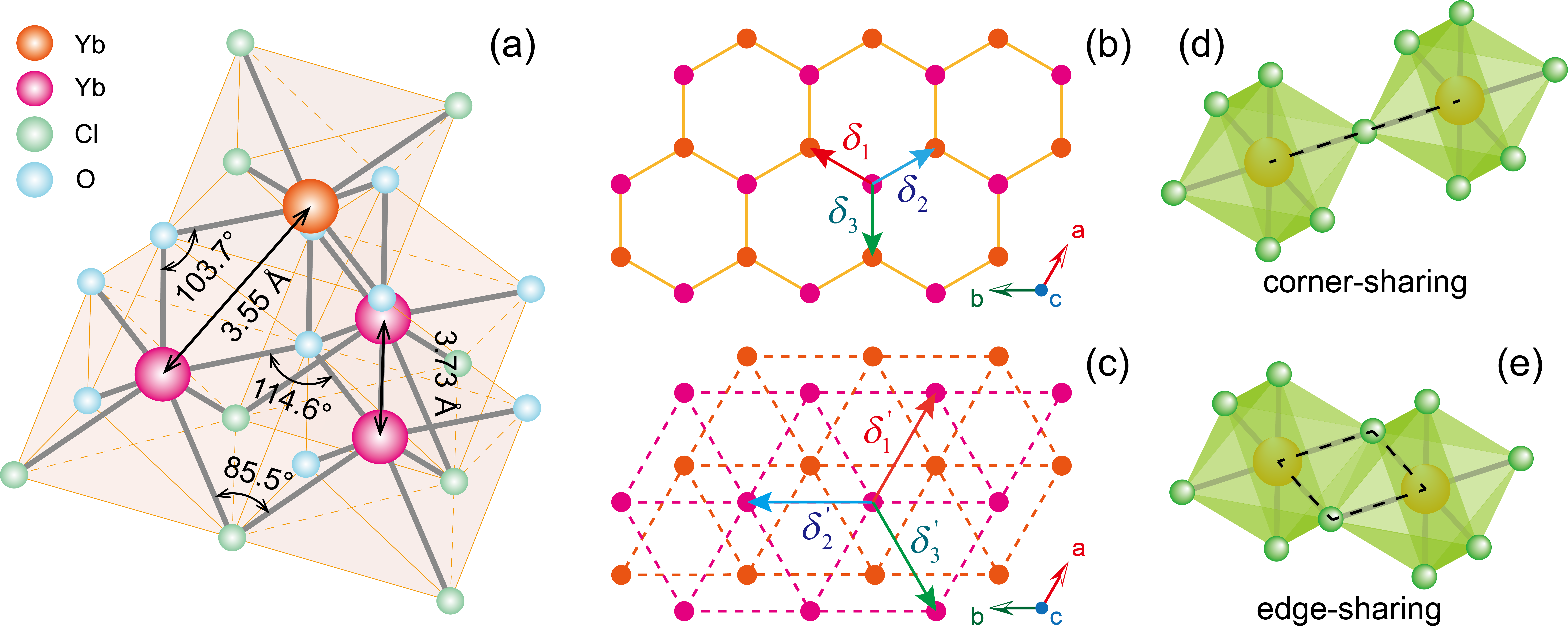}
	\caption{\label{fig:epsart} Local environment around \ce{Yb^{3+}} in \ce{YbOCl}. (a) The octahedral structure formed by \ce{Yb^{3+}} ions and surrounding anions, as well as the bond lengths and bond angles between nearest neighbor (NN) magnetic ions and next nearest neighbor (NNN) magnetic ions. (b) A honeycomb lattice structure formed by NN magnetic ions. (c) A double-layer triangular lattice structure formed by NNN magnetic ions. (d) The schematic of corner-sharing structure. (f) The schematic of edge-sharing structure.}
\end{figure*}

In this paper, we concentrate on the experimental exploration of the spin ground state in \ce{YbOCl}. The magnetization and heat capacity measurements indicate that \ce{YbOCl} undergoes an antiferromagnetic (AFM) transition at 1.3 K. Elastic neutron scattering experiments down to 50 mK have shown magnetic Bragg peaks below 1.3 K and determined the magnetic propagation vector $\vec{Q}$ to be $\left[0, 0, 1.5\right]$. 
A honeycomb-triangle spin Hamiltonian including the nearest neighbor (NN) and next nearest neighbor (NNN) spin interactions has been proposed to accurately describe the spin system. 
The full diagonalization (FD) simulations for magnetization and heat capacity under high magnetic fields determine the spin interactions in the Hamiltonian. The spin interactions are jointly determined by simulating the inelastic neutron scattering spectra (INS) of polycrystalline samples in the ordered phase with linear spin wave theory. The obtained Kitaev interaction is comparable to the Heisenberg one. Using the spin parameters, density matrix renormalization group (DMRG) calculations give an A-type AFM phase, i.e., an AFM ordering along c-axis accompanied by an in-plane ferromagnetic ordering, quantitatively in agreement with neutron and thermodynamic measurements.
The neutron scattering and thermodynamic measurements under magnetic fields allow us to establish a magnetic field-temperature phase (H-T) diagram around the ground state. Interestingly, the AFM order is suppressed by a modest magnetic field of $\sim$ 0.3 T in YbOCl, indicating that the zero-field AFM ground state exhibits notable Kitaev interactions. The application of external magnetic fields, serving as a transverse field term, accentuates quantum fluctuations, suggesting an intricate connection between these fluctuations and the proximity to the Kitaev spin liquid-like behavior.

\begin{figure*}[t]
	\includegraphics[scale=0.9]{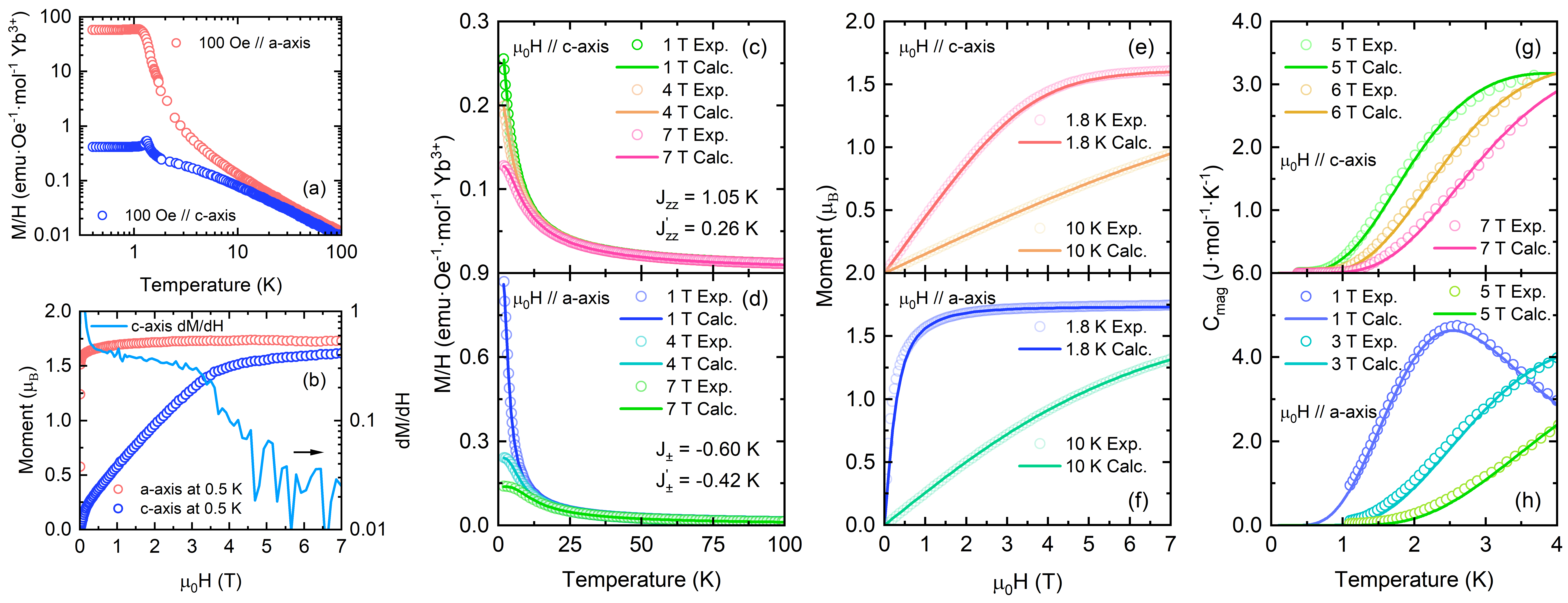}
	\caption{\label{fig:epsart} Magnetization and heat capacity under magnetic fields. (a) and (b) M/H-T and M-H along a-axis (red circles) and along c-axis (blue circles) in \ce{YbOCl} at low temperatures. M/H-T data (open circles) and the simulations (solid lines) along c-axis (c) and a-axis (d) under different magnetic fields. The M-H data along c-axis (open circles) and the simulations (solid lines) (e) and a-axis (f) at 1.8 and 10 K. Magnetic heat capacity under magnetic fields along c-axis (g) and a-axis (h). The open circles and the solid lines represent the experiments and the simulations, respectively.}
\end{figure*}

\subsection{Samples, experiments, and methods}
YbOCl crystals were grown using the anhydrous \ce{YbCl3}-flux method\cite{JiantingJi}.
X-ray diffraction (XRD) and energy dispersive X-ray (EDX) measurements confirmed the high quality of the samples\cite{PhysRevResearch.4.033006}.
Nearly 5 mg of YbOCl single crystals were prepared for thermodynamics measurements. 
The principal axes of single crystals were determined using Laue diffraction plus an optical microscope for further measurements.
The temperature dependent magnetization (M/H-T) and magnetic field dependent magnetization (M-H) measurements along c-axis and a-axis from 0.5 to 2 K under magnetic fields were carried out using a Quantum Design MPMS with a He3 refrigerator.
M/H-T and M-H measurements along c-axis and a-axis from 1.8 to 100 K under magnetic fields were carried out using a Quantum Design PPMS with Vibration Sample Meter (VSM).
Heat capacity and AC susceptibility from 50 mK to 4 K under magnetic fields were measured using a Quantum Design PPMS with a dilution refrigerator (DR).
Elastic neutron scattering was conducted on $\sim$2 g polycrystalline samples down to 50mK\cite{neutron}, and the reflections were determined using FullProf\cite{RODRIGUEZCARVAJAL199355} software.
The magnetic symmetry approach is available on the Bilbao Crystallographic Server\cite{doi:10.1146/annurev-matsci-070214-021008}. 
Approximately 150 single crystal pieces, totaling around 0.2 g of YbOCl\cite{neutron2}, were meticulously co-aligned in the (H0L) scattering plane on copper plates for the neutron experiment to detect the magnetic Bragg peaks.
This experiment was conducted using the cold-neutron Triple-Axis Spectrometer (TASP) at the Swiss Neutron Spallation Source SINQ, located at the Paul Scherrer Institut (PSI) in Switzerland.
The measurements on TASP were completed with a fixed \ce{E_f} = 3.5 meV. A horizontal field cryomagnet, specifically MA02, equipped with a dilution refrigerator, was used to generate a maximum magnetic field of 2 T at temperatures as low as 0.12 K. This setup enabled the application of magnetic fields along c-axis during the experiment.
The 12-spin sites FD was employed to simulate M/H-T, M-H, and heat capacity data under different magnetic fields.
The ground state magnetic structure on a quasi-two-dimensional plane composed of magnetic ions is calculated by the DMRG method\cite{ITensor, ITensor-r0.3}. The PBC was employed to get accurate DMRG results. The clusters used in the DMRG simulations are $L_{x} \times L_{y} = 12 \times 6$ and $L_{x} \times L_{y} = 12 \times 12$. The truncation error was kept below 10$^{-5}$ and we performed 30 sweeps to improve the accuracy of the simulations.
The powder-averaged spin wave calculations along a specified path in the Brillouin zone were based on the SpinW package\cite{Toth_2015}.

\subsection{Model Hamiltonian}
\ce{YbOCl} has a structure of two adjacent triangular magnetic layers. The distance between the NNN magnetic ions ($\sim$ 3.73 \AA) is comparable with that between the NN magnetic ions ($\sim$ 3.55 \AA). We can have a more accurate description on the spin system by taking the NNN spin interactions into account (Fig. 1(a)).
The honeycomb spin lattice formed by the NN magnetic ions (Fig. 1(b)) strictly satisfies the three-fold rotational symmetry $C_{3}$ required by the Kitaev model, while the triangular lattice formed by the NNN magnetic ions (Fig. 1(c)) follows the six-fold rotational symmetry $C_{6}$.
Considering the strong SOC of \ce{Yb^{3+}} and the local octahedral structure, the off-diagonal spin interactions are expected to play an important role. Therefore, the NN spin Hamiltonian can be represented as follows\cite{10.21468/SciPostPhysCore.3.1.004}:
\begin{equation}
	\begin{split}
		\hat H_{Honeycomb} = & \sum\limits_{\langle ij\rangle } {{J_{zz}}} S_i^zS_j^z + {J_ \pm }\left( {S_i^ + S_j^ -  + S_i^ - S_j^ + } \right)\\
		& + {J_{ \pm  \pm }}\left( {{\gamma _{ij}}S_i^ + S_j^ +  + \gamma _{ij}^*S_i^ - S_j^ - } \right)\\
		& + {J_{z \pm }}\left( {{\gamma _{ij}}S_i^ + S_j^z + \gamma _{ij}^*S_i^ - S_j^ z  + \langle i \leftrightarrow j\rangle } \right)
	\end{split}
\end{equation}
where $J_{zz}$, $J_{\pm}$, $J_{\pm\pm}$, and $J_{z\pm}$ are the NN spin-exchange interactions,
$\gamma_{ij}$ corresponds to 1, $e^{2i\pi/3}$, and $e^{-2i\pi/3}$ along the three bonds of $\delta_{1}$, $\delta_{2}$, and $\delta_{3}$ in the honeycomb lattice (Fig. 1(b)).
And the NNN spin Hamiltonian follows\cite{PhysRevLett.115.167203}:
\begin{equation}
	\begin{split}
		\hat H_{Triangular} = & \sum\limits_{\langle\langle ik\rangle\rangle } {{J_{zz}^{'}}} S_i^zS_k^z + {J_{\pm}^{'} }\left( {S_i^ + S_k^ -  + S_i^ - S_k^ + } \right)\\
		& + {J_{ \pm  \pm }^{'}}\left( {{\gamma_{ik}^{'}}S_i^ + S_k^ +  + \gamma _{ik}^{'*}S_i^ - S_k^ - } \right)\\
		& - \frac{i{J_{z \pm }^{'}}}{2}\left( {{\gamma _{ik}^{'*}}S_i^ + S_k^z + \gamma _{ik}^{'}S_i^ - S_k^ z  + \langle i \leftrightarrow k\rangle } \right)
	\end{split}
\end{equation}
where $J_{zz}^{'}$, $J_{\pm}^{'}$, $J_{\pm\pm}^{'}$, and $J_{z\pm}^{'}$ are the NNN spin-exchange interactions,
$\gamma_{ik}^{'}$ corresponds to 1, $e^{2i\pi/3}$, and $e^{-2i\pi/3}$ along the other three bonds of $\delta_{1}^{'}$, $\delta_{2}^{'}$, and $\delta_{3}^{'}$ in the triangular lattice(Fig. 1(c)).
As a higher-order term, Dzyaloshinshii-Moriya (DM) interaction is neglected here. The model Hamiltonian sets a foundation, on which we can carry out quantitative analysis and simulations on the thermodynamic and neutron experiments.

Further discussion can be conducted on the spin Hamiltonian for \ce{YbOCl}.
First, the spin Hamiltonian is equivalent to the J-K-$\Gamma$-$\Gamma$' model\cite{10.21468/SciPostPhysCore.3.1.004, PhysRevX.9.021017, SI}, which allows a mutual conversion of spin exchange parameters between the two models.
The transformation formulas between the spin Hamiltonian and the J-K-$\Gamma$-$\Gamma$' in honeycomb lattice are as follows\cite{10.21468/SciPostPhysCore.3.1.004}:
\begin{equation}
	\begin{array}{l}
		J_{1} = \frac{4}{3}{J_ \pm } - \frac{{2\sqrt 2 }}{3}{J_{z \pm }} - \frac{2}{3}{J_{ \pm  \pm }} + \frac{1}{3}{J_{zz}}\\
		\\
		K_{1} = 2\sqrt 2 {J_{z \pm }} + 2{J_{ \pm  \pm }}\\
		\\
		\Gamma_{1}  =  - \frac{2}{3}{J_ \pm } - \frac{{2\sqrt 2 }}{3}{J_{z \pm }} + \frac{4}{3}{J_{ \pm  \pm }} + \frac{1}{3}{J_{zz}}\\
		\\
		\Gamma_{1}' =  - \frac{2}{3}{J_ \pm } + \frac{{2\sqrt 2 }}{3}{J_{z \pm }} - \frac{4}{3}{J_{ \pm  \pm }} + \frac{1}{3}{J_{zz}}
	\end{array}
\end{equation}
Similarly, the transformation formulas between the spin Hamiltonian and the J-K-$\Gamma$-$\Gamma$' in triangular lattice are as follows\cite{PhysRevX.9.021017}:
\begin{equation}
	\begin{array}{l}
		J_{2} = \frac{1}{3}\left( 2J_{\pm\pm}' + J_{zz}' + 2J_{\pm\pm}' + \sqrt{2}J_{z\pm}' \right)\\
		\\
		K_{2} = -2J_{\pm\pm}' - \sqrt{2}J_{z\pm}'\\
		\\
		\Gamma_{2}  =  \frac{1}{3}\left( -J_{\pm}' + J_{zz}' - 4J_{\pm\pm}' + \sqrt{2}J_{z\pm}' \right)\\
		\\
		\Gamma_{2}' =  \frac{1}{6}\left( -2J_{\pm}' + 2J_{zz}' + 4J_{\pm\pm} - \sqrt{2}J_{z\pm}' \right)
	\end{array}
\end{equation}
From the transformation formulas, it is evident that the Kitaev terms are associated with off-diagonal interaction both in the honeycomb and triangular lattice.
In rare earth magnetic ions, the strong SOC generates significant magnetic anisotropy, which lays the foundation for exploring Kitaev physics in rare earth magnets.
Second, we can discuss the origins of Kitaev interactions from a crystallographic perspective\cite{TREBST20221}.
Two types of configurations involving magnetic ions are frequently observed in various magnetic materials: one is the corner-sharing configuration as shown in Fig. 1(d), and the other is the edge-sharing configuration depicted in Fig. 1(f). For the corner-sharing configuration, of the angle between the magnetic ion-anion-magnetic ion is 180°, the Kitaev interaction K can be completely canceled out. For the edge-sharing configuration, of the angle between the magnetic ion-anion-magnetic ion is 90°, the Heisenberg interaction J can be completely canceled. If the magnetic ions and coordinates ions can form a perfect cubic structure, the off-diagonal terms $\Gamma$ and $\Gamma$' in the J-K-$\Gamma$-$\Gamma$' model vanish completely. In real materials, achieving the ideal crystal structure mentioned above is challenging, hence terms like J, K, $\Gamma$, and $\Gamma$' are present.
Considering these factors, the application of the spin Hamiltonians described by Eq. (1) and (2) is reasonable in \ce{YbOCl}.

\begin{table*}
	\caption{\label{tab:table1} Spin-exchange interactions in model Hamiltonian}
	\begin{ruledtabular}
		\begin{tabular}{cccccccc}
			$J_{zz}$ (K) & $J_{\pm}$ (K) & $J_{\pm\pm}$ (K) & $J_{z\pm}$ (K) & $J_{zz}$' (K) & $J_{\pm}$' (K) & $J_{\pm\pm}$' (K) & $J_{z\pm}$' (K) \\
			\hline
			1.05 & $-0.6$ & 0.39 & 0 & 0.26 & $-0.42$ & 0.13 & 0 \\
		\end{tabular}
	\end{ruledtabular}
\end{table*}

\begin{table*}
	\caption{\label{tab:table2} Spin-exchange interactions in J-K-$\Gamma$-$\Gamma$' model}
	\begin{ruledtabular}
		\begin{tabular}{cccccccc}
			$J_{1}$ (K) & $K_{1}$ (K) & $\Gamma_{1}$ (K) & $\Gamma_{1}$' (K) & $J_{2}$ (K) & $K_{2}$ (K) & $\Gamma_{2}$ (K) & $\Gamma_{2}$' (K) \\
			\hline
			$-0.71$ & 0.78 & 1.27 & 0.49 & $-0.11$ & $-0.26$ & 0.05 & 0.07 \\
		\end{tabular}
	\end{ruledtabular}
\end{table*}

\subsection{Magnetic phase transition of \ce{YbOCl}}
The M/H-T measurements along c-axis and a-axis (Fig. 2(a)) under 0.01 T reveal a transition at 1.3 K. The absence of hysteresis loops in the M-H curve at 0.5 K suggests that it is an AFM transition\cite{SI}. 
A small magnetic field of $\sim$ 0.1 T saturates the magnetic moments along a-axis, while the saturation along c-axis occurs at around 3.5 T (Fig. 2(b)).
The strong anisotropy is consistent with an in-plane ferromagnetic ordering and an AFM ordering along c-axis, which will be clarified later with more neutron experiments and simulations.
The transition is also evidenced by heat capacity measurements from 50 mK to 4 K (Fig. 3(a)). The magnetic entropy has not fully reverted to $Rln2$ up to 4 K\cite{SI} and a higher temperature is required to completely achieve spin disordering. It means that an effective spin-1/2 model remains valid for the spin system below 4 K.

The exchange parameters in the spin Hamiltonian are jointly determined by simulating thermodynamic data under high magnetic fields with FD calculations and INS spectra with linear spin wave theory. 

The diagonal spin interactions $J_{zz}$, $J_{zz}^{'}$, $J_{\pm}$, and $J_{\pm}^{'}$ are first given by the fitting of M/H-T data. At the paramagnetic state and under high magnetic fields, spins are fully polarized and the off-diagonal terms can be ignored. The high-temperature series expansion (HTSE) technique confirms that the effect of the diagonal terms on susceptibility decays as $T^{-2}$ with temperatures while for the off-diagonal ones it decays as $T^{-3}$ in a specified direction\cite{10.21468/SciPostPhysCore.3.1.004, PhysRevLett.115.167203}. This is also verified by our numerical calculations, which show little difference in simulating M/H-T curves with/without the off-diagonal interactions in the low-temperature ($\sim$ 1.8 K) range\cite{SI}.

The M/H-T measurements under magnetic fields along c-axis and a-axis from 1.8 to 100 K are shown in Fig. 2(c) and 2(d), respectively. The 12-spin sites FD simulations of the M/H-T data output an optimal set of parameters, namely $J_{zz}$ = 1.05 K, $J_{zz}^{'}$ = 0.26 K, $J_{\pm}$ = $-0.6$ K, and $J_{\pm}^{'}$ = $-0.42$ K. The parameters allow us to reproduce M/H-T under other magnetic fields\cite{SI} and M-H curves at 1.8 K and 10 K (Fig. 2(e) and 2(f)), which are highly consistent with the experimental results. The square norm of the residuals of $J_{zz}$ and $J_{zz}^{'}$ are 9.306$\times$10$^{-4}$, and 0.011 for $J_{\pm}$ and $J_{\pm}^{'}$. 

Heat capacity generally reflects the energy spectrum of microscopic states. The off-diagonal terms are automatically included in the magnetic heat capacity. Therefore, the off-diagonal terms can be obtained by simulating magnetic heat capacity under high magnetic fields. 

The magnetic moments along c-axis and a-axis at 1.8 K saturate at 5 T and 1 T, respectively. Therefore, magnetic heat capacity measurements under 5 T, 6 T, and 7 T along c-axis and 1 T, 3 T, and 5 T along a-axis are carried out (Fig. 2(g) and 2(h)). The FD simulations give the off-diagonal terms of $J_{\pm\pm}$ = 0.39 K, $J_{\pm\pm}^{'}$ = 0.13 K, $J_{z\pm}$ $\sim$ 0, and $J_{z\pm}^{'}$ $\sim$ 0.
We have summarized these fitted parameters in Table I.
The square norms of the residuals of $J_{\pm\pm}$, $J_{\pm\pm}^{'}$, $J_{z\pm}$ and $J_{z\pm}^{'}$ are 0.77. The DMRG calculations verify that both $J_{z\pm}$ and $J_{z\pm}^{'}$ are close to zero\cite{SI}.

Meanwhile, the INS spectrum down to 50 mK has been measured (Fig. 3(d)) and employed to make a joint determination on the exchange parameters. The exchange parameters given by thermodynamic experiments are applied to the spin Hamiltonian to simulate the powder-averaged INS spectrum at 50 mK with linear spin wave (LSW) theory. The optimized LSW calculations well reproduce the experimental magnetic excitations (Fig. 3(e)). The exchange parameters examined and refined by the LSW simulations are finally pinned down.
We also calculated the dispersion of \ce{YbOCl} along the specified paths (Fig. 3(f)) in the Brillouin zone, as shown in Fig. 3(g).
Calculations reveal no dispersion from the $G$ point to the $A$ point, aligning with the perfect two-dimensional properties of \ce{YbOCl}, which suggest a lack of interlayer spin exchange interactions.
Excitations from the $M$ point through the $G$ point to the $K$ point manifest a distinct V-shape structure, aligning closely with features observed at low $|\vec{Q}|$ in powder INS spectrum.
This offers a valuable reference for future investigations of the INS spectra of \ce{YbOCl} single crystals.

Based on the Eq. (3) and Eq. (4), we can transform spin exchange parameters between the two models, resulting in $J_{1}$ $\sim$ $-0.71$ K, $K_{1}$ $\sim$ 0.78 K, $\Gamma_{1}$ $\sim$ 1.27 K, and $\Gamma_{1}^{'}$ $\sim$ 0.49 K for the honeycomb lattice, and $J_{2}$ $\sim$ $-0.11$ K, $K_{2}$ $\sim$ $-0.26$ K, $\Gamma_{2}$ $\sim$ 0.05 K, and $\Gamma_{2}^{'}$ $\sim$ 0.07 K for the triangular lattice. 
To compare these parameters more clearly, we have summarized them in Table II.

For the $J-K-\Gamma-\Gamma^{'}$ model, it is clearer that the NNN spin interactions are significantly smaller than the NN ones, demonstrating that the spin interactions in the honeycomb lattice play a major role in the magnetism of YbOCl. 
Although the ferromagnetic Heisenberg interactions lead to a magnetically ordered ground state in YbOCl, the antiferromagnetic Kitaev interaction ($K_{1}$ $>$ 0) in \ce{YbOCl} provides the possibility to explore richer Kitaev physics\cite{PhysRevB.83.245104, PhysRevB.97.241110}.
The positive Kitaev term, compared to the negative one in $\alpha$-\ce{RuCl3}\cite{RN25}, is more likely to host stable topological excitations. The ratio $|K_{1}/J_{1}|$ $\sim$ 1.1, comparable to those in $\alpha$-\ce{RuCl3}\cite{RN25,RN27} and \ce{Na2Co2TeO6}\cite{RN28}, indicates that Kitaev spin physics plays an essential role in YbOCl. 
The relatively large off-diagonal terms $\Gamma_{1}$ and $\Gamma_{1}^{'}$ are directly related to the octahedral structure of \ce{YbCl3O4} with distortions.

\begin{figure*}[t]
	\includegraphics[scale=0.85]{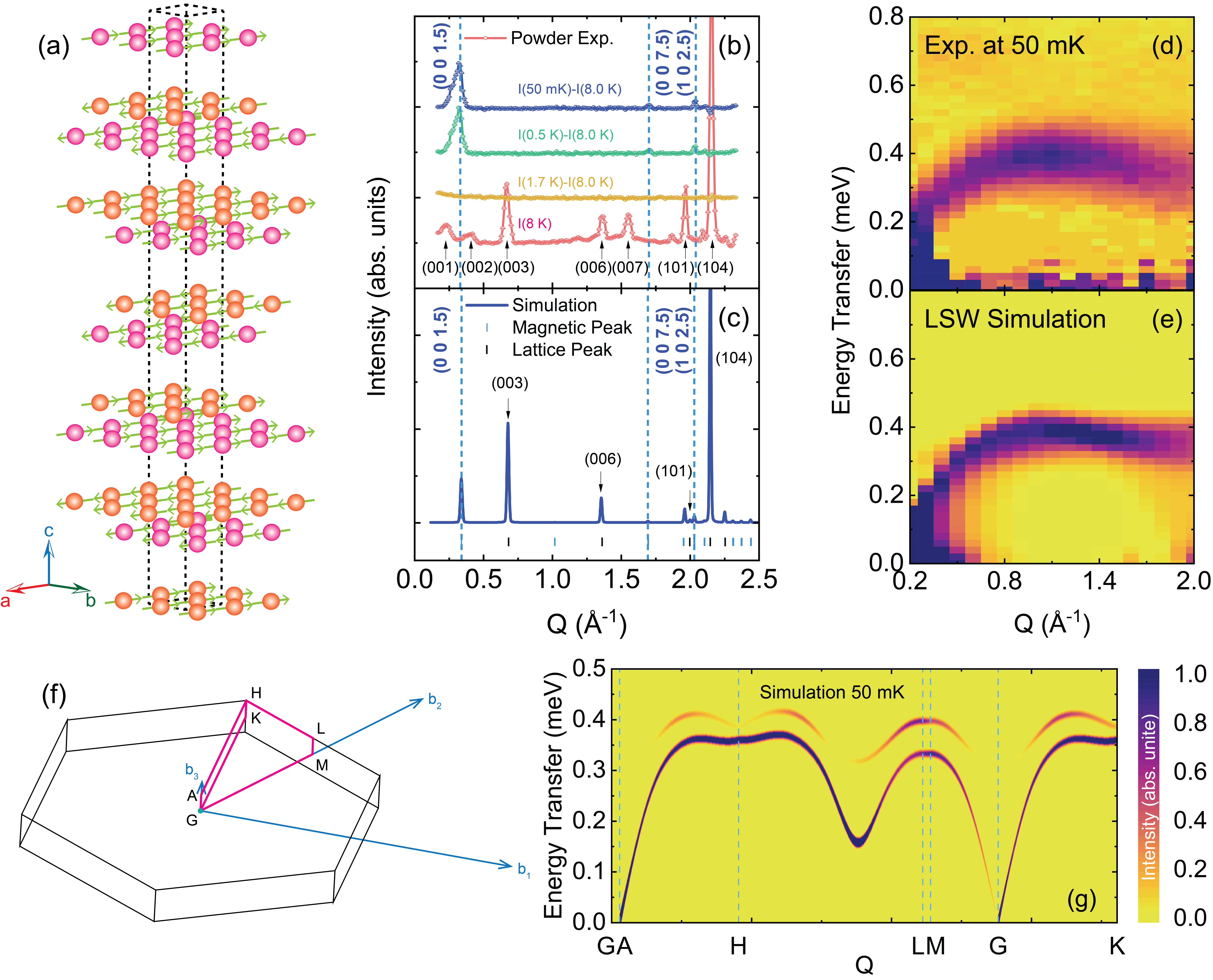}
	\caption{\label{fig:epsart} Ground state magnetic structure and neutron scattering experiments. (a) Three-dimensional ground state magnetic structure of YbOCl with the magnetic propagation vector $\vec{Q}$ = $\left[0, 0, 1.5\right]$. 
		(b) Elastic neutron scattering spectra on polycrystals down to 50 mK. The blue dashed lines mark the magnetic Bragg peaks below the transition temperature 1.3 K.  (c) Elastic neutron scattering spectrum simulated with the magnetic structure of \ce{YbOCl}. The black numbers mark the wave vector of the magnetic peaks, and the red numbers mark the lattice vectors. (d) Experimental INS spectrum on polycrystals at 50 mK. (e) The simulated powder-averaged INS by linear spin wave theory. (f) The Brillouin zone of \ce{YbOCl}. The red solid lines mark the path connecting the high-symmetry points in the Brillouin zone.  (g) The simulated INS spectrum of \ce{YbOCl} based on the LSW theory.}
\end{figure*}

\subsection{Ground state magnetic structure}
We have conducted elastic neutron scattering experiments on polycrystals and down to 50 mK to determine the ground-state magnetic structure of YbOCl. Three new reflections at $|\vec{Q}| = 0.33$ $\AA^{-1}$, $|\vec{Q}| = 1.70$ $\AA^{-1}$, and $|\vec{Q}| = 2.03$ $\AA^{-1}$ in the spin-ordered phase have been highlighted using the spectrum at 8 K as the background (Fig. 3(b)). 
The analysis and fitting based on the FullProf program\cite{RODRIGUEZCARVAJAL199355}, determine that the wave vectors of the magnetic Bragg reflections are $\left[0, 0, 1.5\right]$, $\left[0, 0, 7.5\right]$, and $\left[1, 0, 2.5\right]$, respectively, and the magnetic propagation vector is determined to be $\vec{Q}$ = $\left[0, 0, 1.5\right]$.
Hence, a ferromagnetic magnetic structure forms within ab-plane while an AFM structure develops along c-axis. The three-dimensional ground state magnetic structure is shown in Fig. 3(a).
Combing the magnetization data from Fig. 2(a) and 2(b), we can infer that the in-plane ferromagnetic structure in \ce{YbOCl} primarily arises from anisotropic spin-exchange interactions, whereas the antiferromagnetic alignment along the c-axis is associated with the interlayer van der Waals forces.
Since our focus is on quantum magnetism resulting from spin-exchange interactions, it is reasonable to overlook the interlayer van der Waals forces in the analysis of magnetism of \ce{YbOCl}.

The magnetic structure is also supported by INS experiments at 50 mK (Fig. 3(d)), as mentioned above. And it is further examined by two-dimensional DMRG calculations with the experimentally determined exchange parameters, which demonstrate that the magnetic moments almost lie in ab-plane and form a ferromagnetic ordering along a-axis\cite{SI}.

The in-plane ferromagnetic and interlayer AFM ordering in \ce{YbOCl} is distinct from the observations in some other Kitaev compounds. For instance, the ground-state magnetic structure in $\alpha$-\ce{RuCl3}\cite{PhysRevB.93.134423} and \ce{Na2IrO3}\cite{PhysRevB.85.180403,PhysRevB.83.220403} is a zigzag-ordered state, and the ground state of \ce{YbCl3}\cite{PhysRevB.100.180406,PhysRevB.102.014427,Hao2020} is a Neel one.
The difference is closely related to the fact that $J_{\pm}$ and $J_{\pm}^{'}$ are negative (ferromagnetic) while $J_{zz}$ and $J_{zz}^{'}$ are positive (AFM) in YbOCl.

\begin{figure*}[t]
	\includegraphics[scale=0.8]{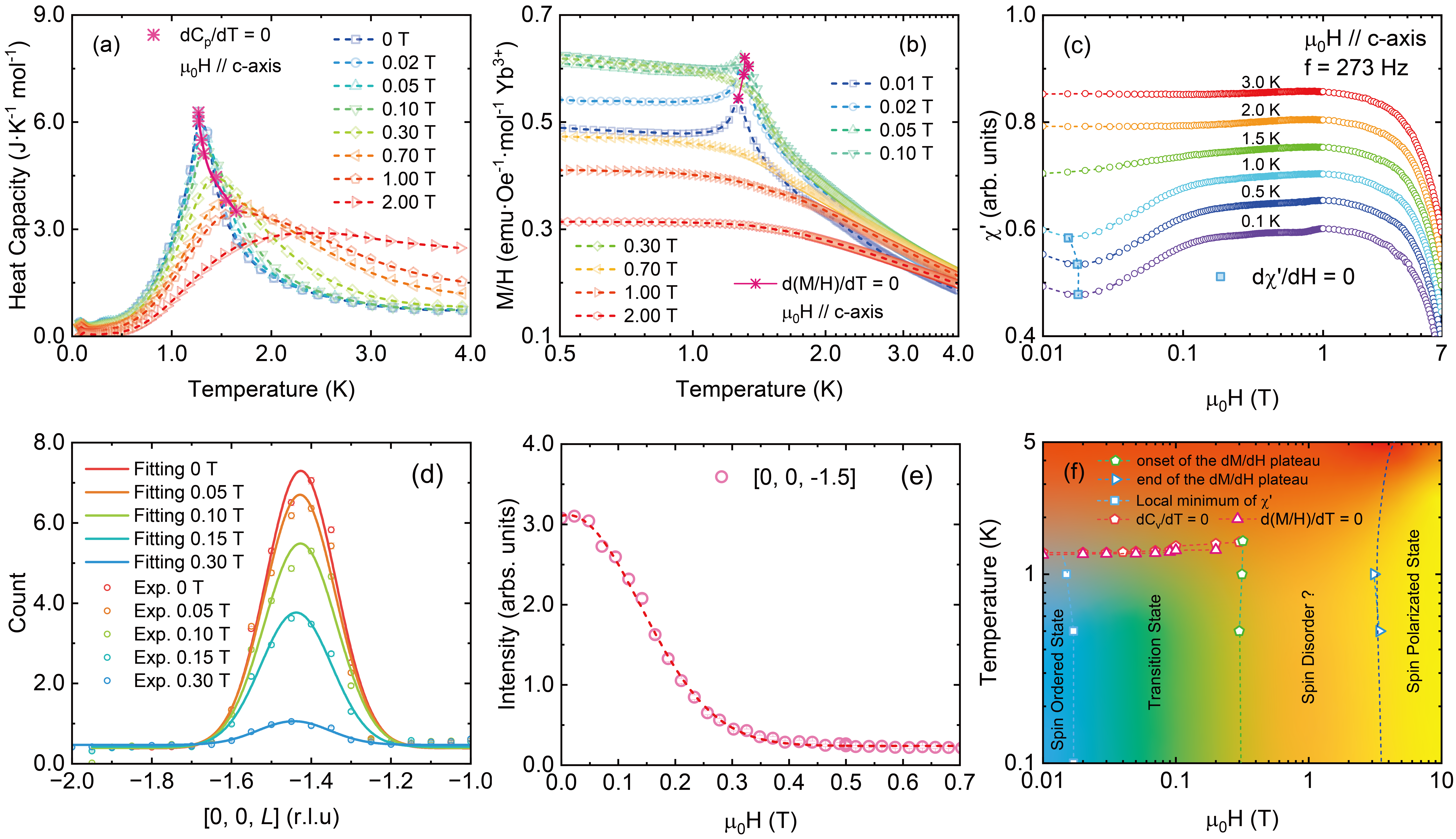}
	\caption{\label{fig:epsart} Magnetic field dependent experiments and H-T phase diagram. (a) Heat capacity under magnetic fields along c-axis. The red stars trace the
		peak change in heat capacity with magnetic fields. (b) M/H-T under magnetic fields along c-axis. The red stars mark the evolution of the phase transition with magnetic fields. More heat capacity and M/H-T data can be found in the supplementary materials\cite{SI}. (c) AC susceptibility $\chi'$ along c-axis at different temperatures. The blue squares mark the local minimum of $\chi'$. (d) The single-crystal neutron diffraction scans in the momentum range of $\vec{Q}$ from $\left[0, 0, -2\right]$ to $\left[0, 0, -1\right]$ with magnetic fields at 0.13 K. The circles represent the experimental results and the solid lines are the fitting ones. (e) The scattering intensity of the magnetic Bragg reflection $\vec{Q}$ = $\left[0, 0, -1.5\right]$ varies with magnetic fields at 0.13 K. (f) The H-T phase diagram is determined by heat capacity, M/H-T, and AC susceptibility $\chi'$ along c-axis. The phase boundaries are marked as the red pentagons and red upper triangles with the criteria of $dC_{p}/dT$ = 0 and the $d(M/H)/dT$ = 0, respectively. The blue square markers indicate the boundaries between the spin ordered phase and the transition state by the local minimum of AC susceptibility $\chi'$. The green pentagons and blue inverted triangle markers respectively denote the boundaries between the possible spin disordered phase and the transition state, and between the possible spin disordered phase and the spin-polarized state.}
\end{figure*}

\subsection{Field effects and H-T phase diagram}
We have investigated the ground state magnetism of \ce{YbOCl} under magnetic fields along c-axis (Fig. 4). The transition is gradually suppressed by increasing magnetic fields (Fig. 4(a) and 4(b)). Elastic neutron scattering provides a clear evolution of the magnetic Bragg peak at $\vec{Q}$ = $\left[0, 0, -1.5\right]$ with magnetic field (Fig. 4(d) and 4(e)).
The magnetic Bragg peak goes down with increasing field, and becomes almost invisible for field  $\ge$ 0.3 T. Hence, a relatively small magnetic field can suppress the magnetic phase transition. This coincides with the plateau observed in the $\chi'$-H (Fig. 4(c)) and dM/dH-H (Fig. 2(b)) curves in the range of 0.3 to 3.1 T.

Interestingly, the transition shifts to higher temperatures with increasing magnetic fields. This has been consistently seen in the heat capacity, M/H-T, and AC susceptibility\cite{SI}, distinct from a conventional AFM transition which usually shifts to lower temperatures with magnetic fields. The unusual magnetic field dependence in \ce{YbOCl} arises from the highly anisotropic spin interactions caused by the strong SOC in \ce{Yb^{3+}} ions. For comparison, $\alpha$-\ce{RuCl3}\cite{RN81,PhysRevB.91.180401,PhysRevLett.125.097203}, \ce{YbCl3}\cite{PhysRevB.102.014427,Hao2020}, and \ce{Na2Co2TeO6}\cite{RN23} demonstrate the evolution of a conventional AFM with magnetic fields. 

The above experiments and analysis allow us to establish the H-T phase diagram (Fig. 4(f)). The H-T phase diagram can be divided into five regions. First of all, the red dashed line marks the paramagnetic state out of other spin phases at higher temperatures. 
Below $\sim$1.3 K, the spin system undergoes four states with increasing magnetic fields, i.e., spin ordered state, transition state, the possible spin disordered phase, and spin polarized state. We have more detailed discussions in the following.

The small dip around 0.02 T in AC susceptibility $\chi$' marks the boundary from the spin ordered state to a transition state.
It stems from the weak interlayer AFM coupling, which can be easily disrupted by an external magnetic field. The transition state covers the range of 0.02 T to 0.3 T. 
In addition to the aforementioned thermodynamic data, single crystal elastic neutron scattering experiments under magnetic field along c-axis also demonstrate that the magnetic Bragg reflection of $\left[0, 0, -1.5\right]$ is gradually suppressed with magnetic fields, suggesting a transition state or crossover before entering into the spin disordered state. 
At higher magnetic fields above $\sim$3.1 T, the system enters into a fully polarized state, with the rapid drop in both AC susceptibility $\chi$ (Fig. 4(c)) and dM/dH-H (Fig. 2(b)).

The possible spin disordered phase in the range of 0.3 to 3.1 T is of particular interest. In the region, the transition peaks in heat capacity and M/H-T fade away with magnetic fields and there appears a clear plateau in ac susceptibility (Fig. 4(c)). Generally, AC magnetic susceptibility can be regarded as a spin "response" function and acts as a good indicator of spin fluctuations in a spin system\cite{Topping_2019}.
The formula for the definition of AC magnetic susceptibility is as follow:
\begin{equation}
		\chi' = \lim_{\mu_{0}H \rightarrow 0} \frac{\partial \langle M \rangle}{\partial \mu_{0} H} = \frac{\langle M^{2} \rangle - \langle M \rangle ^{2}}{k_{B}T}
\end{equation}
The plateau-like feature in AC susceptibility $\chi$' corresponds to a region in which spin fluctuations are enhanced by magnetic fields. The feature is verified by the derivatives of M-H curve (Fig. 2(b)). More crucially, the nearly complete suppression of the magnetic Bragg at $\vec{Q}$ = $\left[0, 0, -1.5\right]$ by a magnetic field above 0.3 T, clearly indicates that the AFM order is suppressed by magnetic fields. 
The region of enhanced spin fluctuations may potentially represent a KLS-like state.
If the spins were merely aligning along c-axis with increasing magnetic fields, the AC susceptibility $\chi$' would not exhibit a sustained robust plateau between 0.3 and 3.1 T.
On the other hand, we consider it unlikely that the magnetic moments gradually polarize along the c-axis direction within the magnetic field range of 0.3 to 3 T. If we follow this perspective, the magnetic Bragg peak at $\vec{Q}$ = $\left[0, 0, 1.5\right]$ should not completely disappear before the spins are fully polarized. Conversely, elastic neutron scattering data demonstrate that the magnetic Bragg peak at $\vec{Q}$ = $\left[0, 0, 1.5\right]$ vanished at fields exceeding 0.3 T, which is markedly different from the saturation field of 4 T.
We also estimate the Zeeman energy scale using the experimentally determined parameters. The Zeeman term is $\mu_{0} \mu_{B} g_{c} H_{c}$ ($\mu_{B}$ is Boltzmann's constant, $g_{c}$ $\sim$ 3.2 is Lande g-factor along c-axis, and $H_{c}$ is magnetic field), and a magnetic field of 0.3 T is about 0.6 K. For comparison, the spin interaction is approximately 1.0 K. The energy scale contributed by the magnetic field is insufficient to drive the spins into a polarized state.

The unusual behavior seems related to the fact that the magnetic moment in ab-plane is significantly larger than that along c-axis. Therefore, the magnetic field acts as an effective transverse field along c-axis, which essentially enhances quantum fluctuations and suppresses the magnetically ordered state, along with the off-diagonal spin interactions.
The similar magnetic field-dependent behavior has been witnessed in other KSL candidates, such as $\alpha$-\ce{RuCl3}\cite{PhysRevB.95.180411,PhysRevB.96.041405,PhysRevLett.119.037201,PhysRevLett.119.227208} and \ce{Na2Co2TeO6}\cite{RN23,PhysRevB.94.214416,RN24,PhysRevB.95.094424}. All of them show a spin-ordered ground state at 0 T and are tuned into a KSL-like state by applying magnetic fields. It should be noted that the critical magnetic field is $\sim$ 7 T for $\alpha$-\ce{RuCl3}, $\sim$ 7.5 T for \ce{Na2Co2TeO6}, and only $\sim$ 0.3 T in \ce{YbOCl}.
The greatly reduced critical field in \ce{YbOCl} stems from the strong screening for 4f electrons by the outer 5s and 5p shells. This offers a big advantage of easily manipulating a potential KSL phase, making \ce{YbOCl} a unique and promising candidate for further exploration in this intriguing area of research.

The emergence of the KSL-like state in \ce{YbOCl} is accompanied by several quantum phase transitions and quantum critical points, i.e., from the spin-ordered state to the KSL-like state and from the KSL-like state to the fully polarized state. The unusual KSL-like phase itself and the quantum phase transitions are expected to offer rich Kitaev spin physics which highly requires deep theoretical and experimental explorations in the future.

\emph{Summary} ---
In this study, we experimentally explore the magnetic ground state in the newly reported layered honeycomb antiferromagnet \ce{YbOCl} and discover a possible spin liquid phase. 
By combining thermodynamic and neutron measurements with sophisticated numerical simulations, the fundamental diagonal and off-diagonal exchange coupling parameters for the highly anisotropic spin system are determined, which quantitatively pins down the spin Hamiltonian.
The compound undergoes an AFM transition at 1.3 K. The neutron scattering experiments down to 50 mK and DMRG and spin wave calculations consistently point to the magnetic ground state with a magnetic propagation vector $\vec{Q}$ = $\left[0, 0, 1.5\right]$.
The experiments allow us to establish an H-T phase diagram. Most interestingly, the ordered ground state can be pushed into a spin disordered state with enhanced quantum spin fluctuations (or a possible KSL phase) using a small magnetic field ($\sim$ 0.3 T). 
The present study demonstrates that the van der Waals layered family offers an inspiring playground for anisotropic spin systems, particularly for Kitaev spin physics and potential applications.

\emph{Acknowledgment} ---
This work was supported by the National Key Research and Development Program of China (Grant No. 2022YFA1402700), the National Science Foundation of China (Grant No. 12274186),  the Strategic Priority Research Program of the Chinese Academy of Sciences (Grant No. XDB33010100), and the Synergetic Extreme Condition User Facility (SECUF). A portion of this work was performed on the Steady High Magnetic Field Facility, High Magnetic Field Laboratory, CAS.


\begin{thebibliography}{45}
\bibitem [{neu({\natexlab{a}})}]{neutron}%
	\BibitemOpen
	\href@noop {} {\enquote {\bibinfo {title} {The neutron scattering data were
				collected at the bragg institute of the australian nuclear science and
				technology organisation (ansto) using the pelican time-of-flight
				spectrometer.}}\ } ({\natexlab{a}})\BibitemShut {NoStop}%
\bibitem [{neu({\natexlab{b}})}]{neutron2}%
	\BibitemOpen
	\href@noop {} {\enquote {\bibinfo {title} {The neutron scattering data on
				single crystals were collected on the cold-neutron triple-axis spectrometer
				(tasp) at the swiss neutron spallation source sinq, paul scherrer institut
				(psi), switzerland.}}\ } ({\natexlab{b}})\BibitemShut {NoStop}%
\end{thebibliography}

%

\end{document}